\documentstyle{article}\input amssym.def\input amssym
\newcommand{\no}{\noindent}\textheight=24.8cm\topmargin=-1.2cm\textwidth=15.3cm
\begin{document}\bibliographystyle{plain}
\begin{titlepage}\renewcommand{\thefootnote}{\fnsymbol{footnote}}\large
\hfill\begin{tabular}{l}HEPHY-PUB
701/98\\UWThPh-1998-53\\hep-ph/9812368\\September 1998\end{tabular}\\[3cm]
\begin{center}
{\huge\bf SEMIRELATIVISTIC TREATMENT}\\[2ex]{\huge\bf OF BOUND STATES}\\
\vspace{2cm}{\Large\bf Wolfgang LUCHA\footnote[1]{\normalsize\ {\em E-mail\/}:
v2032dac@vm.univie.ac.at}}\\[.5cm] Institut f\"ur Hochenergiephysik,\\
\"Osterreichische Akademie der Wissenschaften,\\Nikolsdorfergasse 18, A-1050
Wien, Austria\\[1.7cm] {\Large\bf Franz F.
SCH\"OBERL\footnote[2]{\normalsize\ {\em E-mail\/}:
franz.schoeberl@univie.ac.at}}\\[.5cm] Institut f\"ur Theoretische Physik,\\
Universit\"at Wien,\\ Boltzmanngasse 5, A-1090 Wien, Austria\\[2cm]\end{center}
{\em Invited talk by W. Lucha at the Symposium on ``Quarks in Hadrons and
Nuclei'' (202. WE-Heraeus Seminar), September 14 -- 19, 1998, Rothenfels
Castle, Oberw\"olz,~Austria\/}\vfill\begin{center}{\bf Abstract}\end{center}
\normalsize This talk reviews several aspects of the ``semirelativistic''
description of bound states by the spinless Salpeter equation (which
represents the simplest equation of motion incorporating relativistic
effects) and, in particular, presents or recalls some very simple and
elementary methods which allow to derive rigorous statements on the
corresponding solutions, that is, on energy levels as well as wave functions.

\vspace*{6ex}

\normalsize\renewcommand{\thefootnote}{\arabic{footnote}}\end{titlepage}

\section{SPINLESS SALPETER EQUATION}The conceptually simplest bound-state
wave equation incorporating to some extent relativistic effects is the
``spinless Salpeter equation.'' The origin of this equation of motion is by
no means some mystery. Rather, this spinless Salpeter equation has to be
regarded as a well-defined standard approximation~to the Bethe--Salpeter
formalism.

\subsection{Bethe--Salpeter (BS) Formalism}The appropriate framework for the
description of bound states within some relativistic quantum~field theory is
the Bethe--Salpeter formalism. Consider the bound state $|{\rm M}\rangle$
(with momentum $K$ and~energy $E$) of, say, (spin-$\frac{1}{2}$) fermion and
antifermion (with masses $M_1$, $M_2$ and momenta $P_1$, $P_2$,
respectively). Within the BS formalism this bound state is represented by its
{\em Bethe--Salpeter amplitude\/} $\Psi$ which, in momentum space, is defined
as Fourier transform of the time-ordered product of the field operators~of
the two bound-state constituents between the vacuum $|0\rangle$ and the bound
state, after factorizing off~the center-of-momentum motion:
$$\fbox{$\Psi(P)=\exp({\rm i}\,K\,X_{\rm CM})\displaystyle\int{\rm
d}^4X\,\exp({\rm i}\,P\,X)\,\langle 0|{\rm
T}(\psi_1(X_1)\,\bar\psi_2(X_2))|{\rm M}(K)\rangle$}\ ,$$ with the usual
kinematical notions of total momentum $K=P_1+P_2$, relative momentum $P$,
center-of-momentum coordinate $X_{\rm CM}$, and relative coordinate $X\equiv
X_1-X_2$ of the bound-state constituents.

The BS amplitude $\Psi$ satisfies the {\em Bethe--Salpeter equation\/}
\cite{Salpeter51} $$\fbox{$\displaystyle
S_1^{-1}(P_1)\,\Psi(P)\,S_2^{-1}(-P_2)=\frac{{\rm i}}{(2\pi)^4}\int{\rm
d}^4Q\,{\cal K}(P,Q)\,\Psi(Q)$}\ ,$$ which involves two dynamical
ingredients, namely,\begin{itemize}\item on the one hand, the full fermion
propagator of particle $i$ ($i=1,2$), denoted by $S(P,M)$ for~any particle of
momentum $P$ and mass $M$, which, however, is usually approximated by its
free~form $$S_0^{-1}(P,M)=-{\rm i}\,(\gamma_\mu\,P^\mu-M)\ ,$$ with $M$
interpreted as some effective (``constituent'') mass and the propagator $S_0$
understood~as an effective one, and,\item on the other hand, the {\em BS
interaction kernel\/} ${\cal K}(P,Q)$, which is defined (only
perturbatively!)~as sum of all two-particle (``BS-'') irreducible Feynman
diagrams~for two-particle into two-particle scattering (exhibiting,
consequently, a certain amount of gauge dependence), and which is~given, for
instance, in lowest order quantum electrodynamics from one-photon exchange
(in Feynman gauge) by $${\cal
K}(P,Q)=\frac{e^2}{(P-Q)^2}\,\gamma^\mu\otimes\gamma_\mu\ .$$\end{itemize}

By construction, the BS equation is formally exact. Its actual application,
however, faces several, rather well-known {\bf
problems}:\footnote{\normalsize\ The appearance of states with negative or
vanishing norm among the obtained solutions as~well~as questions of
interpretation, for instance, cause additional troubles when dealing with the
BS equation.}\begin{itemize}\item There is no means to compute the BS kernel
beyond the very tight limits of perturbation~theory.\item Even with the BS
kernel at one's disposal, it is, in general, not possible to find the exact
solutions of the BS equation (except for rather few special cases, like the
famous Wick--Cutkosky model describing the interaction of two scalar
particles by exchange of some massless scalar~particle).\item In non-Abelian
gauge theories like, e.g., quantum chromodynamics, since the Dyson--Schwinger
equations connect the propagators and the interaction kernel, the
simultaneous assumption of free propagators and a confining BS kernel is
obviously inconsistent.\end{itemize}

The most straightforward way out is the {\bf reduction} of the
Bethe--Salpeter equation by a series~of approximations:\begin{enumerate}\item
Eliminate any dependence on timelike variables. This (innocent) enterprise
involves two steps:\begin{itemize}\item Adhere to the {\em static
approximation\/} to the BS kernel (which has to be justified {\em a
posteriori\/}), that is, assume that the BS kernel depends only on the
relative three-momenta ${\bf Q}$ and ${\bf P}$ of initial and final state,
respectively: $${\cal K}(P,Q)={\cal K}({\bf P},{\bf Q})\ .$$ (This amounts to
ignoring all retardation effects by assuming instantaneous
interactions.)\item Define the {\em equal-time wave function\/} $\Phi$,
sometimes also called the ``Salpeter amplitude,''~(in momentum space) by
integrating the BS amplitude $\Psi(P)$ over the zero component of the
four-momentum $P$: $$\Phi({\bf P})\equiv\int{\rm d}P_0\,\Psi({\bf P},P_0)\
.$$\end{itemize} This leads to the {\em Salpeter equation\/}
\cite{Salpeter52} $$\Phi({\bf P})=\int\frac{{\rm
d}^3Q}{(2\pi)^3}\left[\frac{\Lambda_1^+\,\gamma_0\,{\cal K}({\bf P},{\bf
Q})\,\Phi({\bf Q})\,\gamma_0\left(\Lambda_2^+\right)^{\rm
c}}{E-\displaystyle\sqrt{{\bf P}_1^2+M_1^2}-\displaystyle\sqrt{{\bf
P}_2^2+M_2^2}}-\frac{\Lambda_1^-\,\gamma_0\,{\cal K}({\bf P},{\bf
Q})\,\Phi({\bf Q})\,\gamma_0\left(\Lambda_2^-\right)^{\rm
c}}{E+\displaystyle\sqrt{{\bf P}_1^2+M_1^2}+\displaystyle\sqrt{{\bf
P}_2^2+M_2^2}}\right],$$ with the energy projection operators for positive or
negative energy of particle $i$ ($i=1,2$)~given, as usual, by
$$\Lambda_i^\pm\equiv\frac{\displaystyle\sqrt{{\bf
P}_i^2+M_i^2}\pm\gamma_0\,(\mbox{\boldmath{$\gamma$}}\cdot{\bf
P}_i+M_i)}{2\,\displaystyle\sqrt{{\bf P}_i^2+M_i^2}}$$ and
$$\left(\Lambda_i^\pm\right)^{\rm
c}\equiv\left(C^{-1}\,\Lambda_i^\pm\,C\right)^{\rm T}={\Lambda_i^\mp}\ ,$$
where $C$ is the charge conjugation matrix. The Salpeter equation is the
equation of motion for the equal-time wave function $\Phi({\bf P})$, with
full relativistic kinematics but in static approximation for the interaction
kernel ${\cal K}$.\item Neglect the second term on the right-hand side of the
Salpeter equation on the basis of the---at least for ``heavy'' bound-state
constituents reasonable---assumption that the denominator~of~the first term
is much smaller than the denominator of the second term: $$E-\sqrt{{\bf
P}_1^2+M_1^2}-\sqrt{{\bf P}_2^2+M_2^2}\ll E+\sqrt{{\bf
P}_1^2+M_1^2}+\sqrt{{\bf P}_2^2+M_2^2}\ .$$ This leads to the {\em reduced
Salpeter equation\/} $$\fbox{$\displaystyle\left(E-\sqrt{{\bf
P}_1^2+M_1^2}-\sqrt{{\bf P}_2^2+M_2^2}\right)\Phi({\bf
P})=\displaystyle\int\frac{{\rm
d}^3Q}{(2\pi)^3}\,\Lambda_1^+\,\gamma_0\,{\cal K}({\bf P},{\bf Q})\,\Phi({\bf
Q})\,\gamma_0\left(\Lambda_2^+\right)^{\rm c}$}\ .$$\item Neglect any
reference to the spin degrees of freedom of the involved bound-state
constituents.\item Assume that the BS kernel ${\cal K}({\bf P},{\bf Q})$
depends only on the difference of the relative momenta~${\bf P}$ and ${\bf
Q}$, which means that the BS kernel is of convolution type (as is, in fact,
frequently the~case): $${\cal K}({\bf P},{\bf Q})={\cal K}({\bf P}-{\bf Q})\
.$$\item Restrict the whole formalism exclusively to positive-energy
solutions, which will be denoted by $\psi$.\end{enumerate} After applying all
these simplifying assumptions and approximations to the BS equation, one
finally ends up with the {\em spinless Salpeter equation\/}
$$\fbox{$\left[\displaystyle\sqrt{{\bf P}_1^2+M_1^2}+\displaystyle\sqrt{{\bf
P}_2^2+M_2^2}+V({\bf X})\right]\psi=E\,\psi$}\ ,$$ involving an interaction
potential, $V({\bf X})$, arising as the Fourier transform of the BS kernel
${\cal K}({\bf P}-{\bf Q})$. In particular, in the center-of-momentum frame
of the two bound-state constituents (i.e., for ${\bf K}={\bf 0}$), this
equation reads $$H|\psi\rangle=E|\psi\rangle\ ,$$ with the Hamiltonian
$$H=\sqrt{{\bf P}^2+M_1^2}+\sqrt{{\bf P}^2+M_2^2}+V({\bf X})\ .$$

\subsection{Equal-Mass Case}

For the special case of equal masses of the two involved bound-state
constituents, i.e., when assuming $$M_1=M_2=M\ ,$$ it is possible to recast
the two-particle spinless Salpeter equation into the equivalent
one-particle~form. In order to see this, consider the semirelativistic
Hamiltonian $H$ for the two-particle spinless Salpeter equation with an
interaction described by a central potential $V(|{\bf X}|)$ of, for instance,
power-law form: $$H=2\,\sqrt{{\bf P}^2+M^2}+\sum_{n\in\Bbb Z}k_n\,R^n\ ,\quad
R\equiv|{\bf X}|\ .$$ One may always perform a scale transformation of the
phase-space variables ${\bf X}$, ${\bf P}$ by some arbitrary scale factor
$\lambda$, $${\bf p}=\lambda\,{\bf P}\ ,\quad{\bf x}=\frac{{\bf X}}{\lambda}\
,$$ since any transformation of this kind (necessarily) preserves the
fundamental commutation relations: $$[{\bf x},{\bf p}]=[{\bf X},{\bf P}]\ .$$
Now,\begin{itemize}\item fix the scale parameter $\lambda$ to the value
$\lambda=2$, which implies for the semirelativistic Hamiltonian
\begin{eqnarray*}H&=&\sqrt{4\,{\bf P}^2+4\,M^2}+\sum_{n\in\Bbb Z}k_n\,R^n\\
&=&\sqrt{{\bf p}^2+4\,M^2}+\sum_{n\in\Bbb Z}k_n\,2^n\,r^n\ ,\end{eqnarray*}
\item identify the mass and coupling-strength parameters for one- and
two-particle form according~to\begin{eqnarray*}m&=&2\,M\
,\\[1ex]a_n&=&2^n\,k_n\ ,\quad n\in\Bbb Z\ ,\end{eqnarray*}\end{itemize} and
arrive---presumably without great surprise---at the equivalent one-particle
Hamiltonian $$H=\sqrt{{\bf p}^2+m^2}+\sum_{n\in\Bbb Z}a_n\,r^n\ ,\quad
r\equiv|{\bf x}|\ .$$

\subsection{(One-Particle) Spinless Salpeter Equation}

In view of the above observation, it is sufficient to confine the present
discussion to the consideration~of some self-adjoint Hamiltonian $H$ of the
form \begin{equation}H=T+V\ ,\label{Eq:SRH}\end{equation} where $T$ denotes
the ``square-root'' operator of the relativistic expression for the free
(kinetic) energy of some particle of mass $m$ and momentum ${\bf p}$,
$$T=T({\bf p})\equiv\sqrt{{\bf p}^2+m^2}\ ,$$ and $V=V({\bf x})$ represents
some arbitrary coordinate-dependent, static interaction-potential operator.
The {\bf spinless Salpeter equation} is nothing else but the eigenvalue
equation for the Hamiltonian~$H$, $$H|\chi_k\rangle=E_k|\chi_k\rangle\ ,\quad
k=0,1,2,\dots\ ,$$ for the (complete set of) Hilbert-space eigenvectors
$|\chi_k\rangle$ of $H$ corresponding to the energy eigenvalues
$$E_k\equiv\frac{\langle\chi_k|H|\chi_k\rangle}{\langle\chi_k|\chi_k\rangle}\
.$$ As such, it represents the simplest relativistic generalization of the
Schr\"odinger equation of standard nonrelativistic quantum theory. {\bf N.B.}
The semirelativistic Hamiltonian $H$ is a {\em nonlocal\/} operator,~i.e.,
\begin{itemize}\item either the relativistic kinetic-energy operator $T$ {\em
in configuration space\/}\item or, in general, the interaction-potential
operator $V$ {\em in momentum space\/}\end{itemize} is nonlocal. Because of
this nonlocality it is somewhat difficult to obtain rigorous {\em analytic\/}
statements on the solutions of this equation of motion.

\section{RELATIVISTIC COULOMB PROBLEM}

Of particular importance in many realms of physics is the (spherically
symmetric) Coulomb potential, with interaction strength parametrized by some
dimensionless coupling (``fine structure'') constant~$\alpha$:
\begin{equation}V({\bf x})=V_{\rm C}(r)=-\frac{\alpha}{r}\ ,\quad
r\equiv|{\bf x}|\ ,\quad\alpha>0\ .\label{Eq:CP}\end{equation}

\subsection{Spinless Relativistic Coulomb Problem}

Since we will use the Coulomb potential in order to illustrate the general
statements derived below,~let us briefly summarize, in roughly chronological
order, the most important knowledge gained until now:\begin{itemize}\item By
examination of the spectral properties of the semirelativistic Hamiltonian
$H$ with a Coulomb interaction potential $V_{\rm C}(r)$, one is able to prove
\cite{Herbst} for this operator $H$\begin{itemize}\item its {\em essential
self-adjointness\/} (which means that the closure of $H$ is self-adjoint) for
$\alpha\le\displaystyle\frac{1}{2}$;\item the existence of its {\em
Friedrichs extension\/} up to a critical value $\alpha_{\rm c}$ of the
coupling constant~$\alpha$: $$\alpha_{\rm c}=\frac{2}{\pi}\ ;$$\item a strict
{\em lower bound\/} on the ground-state energy $E_0$ (that is, on the
operator $H$), given by $$\fbox{$\displaystyle E_0\ge
m\,\sqrt{1-\left(\frac{\pi\,\alpha}{2}\right)^2}\quad\mbox{for}\
\alpha<\frac{2}{\pi}$}\ .$$\end{itemize} Loosely speaking, the Coulombic
Hamiltonian $H$ may be regarded as reasonable operator up~to the critical
coupling constant $\alpha_{\rm c}$.\item For part of the allowed range of the
coupling constant $\alpha$, an improved lower bound can be~found
\cite{Martin}: $$E_0\ge
m\,\sqrt{\frac{1+\sqrt{1-4\,\alpha^2}}{2}}\quad\mbox{for}\
\alpha<\frac{1}{2}\ .$$\item The analytic solution for the corresponding wave
function $\psi$ has been constructed \cite{Durand}; inspecting its behaviour
near the origin, one observes for, e.g., vanishing orbital angular momentum
($\ell=0$) that the configuration-space wave function $\psi({\bf x})$
diverges for small values of $r\equiv|{\bf x}|$ exactly~like $$\lim_{r\to
0}\psi({\bf x})\propto r^{-\nu}\quad\mbox{with}\quad
\nu=\frac{2\,\alpha}{\pi}\left(1+\frac{2\,\alpha}{\pi}+\cdots\right).$$\item
Given the relative uselessness of Temple's inequality, lower bounds on the
spectrum of $H$ may~be found (in a numerical manner) with the help of the
{\bf generalized ``local-energy''~theorem}~\cite{Raynal}:\begin{itemize}\item
Assume\begin{enumerate}\item that the Fourier transform $\widetilde V({\bf
p})$ of the interaction potential $V({\bf x})$ is strictly negative, except
at infinity, as is (certainly) the case for the (attractive) Coulomb
potential (\ref{Eq:CP}),\item that the spectrum of the operator $H$ is
discrete, and\item that the ground state of the Hamiltonian $H$
exists.\end{enumerate}\item Define the ``local energy'' $${\cal E}({\bf
p})\equiv T({\bf p})+\frac{\displaystyle\int{\rm d}^3q\,\widetilde V({\bf
p}-{\bf q})\,\phi({\bf q})} {\phi({\bf p})}\ ,$$ where $\phi({\bf p})$
denotes some suitably chosen, positive trial function, $$\phi({\bf p})>0\
.$$\end{itemize} Then the lowest-lying eigenvalue $E_0$ of the Hamiltonian
$H$ is bounded from below and above~by\footnote{\normalsize\ The lower bound
even holds if the assumption on the discreteness of the spectrum is not
fulfilled.} $$\fbox{$\displaystyle\inf_{{\bf p}}{\cal E}({\bf p})\le
E_0<\sup_{{\bf p}}{\cal E}({\bf p})$}\ .$$\end{itemize}

\newpage

\vspace*{-7ex}
\begin{table}[h]\caption[]{Numerically (!) obtained \cite{Raynal} lower and
upper bounds on the ground-state energy $E_0(\alpha)$}\label{Tab:Nolaub}
\begin{center}\begin{tabular}{lll}\hline\hline&&\\[-1.5ex]
\multicolumn{1}{c}{$\alpha$}&\multicolumn{2}{c}
{$\displaystyle\frac{E_0}{m}$}\\[2ex]
\multicolumn{1}{c}{}&\multicolumn{1}{c}{$\quad$Lower Bound$\quad$}&
\multicolumn{1}{c}{$\quad$Upper Bound$\quad$}\\[1ex]\hline&&\\[-1.5ex]
$\quad$0.0155522$\quad$&$\qquad$0.9998785&$\qquad$0.9998791\\
$\quad$0.1425460$\quad$&$\qquad$0.989458&$\qquad$0.989613\\
$\quad$0.2599358$\quad$&$\qquad$0.96309&$\qquad$0.96364\\
$\quad$0.3566678$\quad$&$\qquad$0.92578&$\qquad$0.92673\\
$\quad$0.4359255$\quad$&$\qquad$0.88013&$\qquad$0.88139\\
$\quad$0.5$\quad$&$\qquad$0.82758&$\qquad$0.82910\\
$\quad$0.5505528$\quad$&$\qquad$0.76908&$\qquad$0.77075\\
$\quad$0.5887832$\quad$&$\qquad$0.7050&$\qquad$0.7069\\
$\quad$0.6155367$\quad$&$\qquad$0.6359&$\qquad$0.6379\\
$\quad$0.6313752$\quad$&$\qquad$0.5616&$\qquad$0.5637\\
$\quad$0.6366198$\quad$&$\qquad$0.4825&$\qquad$0.4843\\[1ex]
\hline\hline\end{tabular}\end{center}\end{table}

\vspace*{-3.5ex}

\no
With the help of this local-energy theorem, the ground-state energy,
considered as a function $E_0(\alpha)$~of the coupling strength $\alpha$, has
been restricted to some remarkably narrow band (Table~\ref{Tab:Nolaub}). In
particular, at the critical coupling constant $\alpha_{\rm c}$ one finds
$$0.4825\le\frac{E_0(\alpha=\alpha_{\rm c})}{m}\le 0.4842910\quad\mbox{for}\
\alpha=\alpha_{\rm c}\ .$$ From this one learns that the ground-state energy
$E_0$ at the critical coupling constant $\alpha_{\rm c}$ is definitely
nonvanishing (which is evidently {\em not\/} clear from the lower bounds
quoted above).

\subsection{Power Series Expansions}

The systematic series expansions of the energy eigenvalues $E(n_{\rm r},\ell)$
in powers of the coupling constant $\alpha$ (up to and including the
order~$O(\alpha^7)$) for states with radial quantum number $n_{\rm r}$ and
orbital~angular momentum $\ell$ have been worked out in different ways
\cite{LeYaouanc,Brambilla}. The {\bf disadvantage} of these power-series
expansions is that they are only significant for a region of rather small
values of the coupling constant $\alpha$. These expansions read, for
instance,\begin{itemize}\item for $n_{\rm r}=0$, $\ell=0$ (i.e., for the
ground state):\begin{eqnarray*}
\frac{E(n_{\rm r}=0,\ell=0)}{m}&=&1-\frac{\alpha^2}{2}-\frac{5\,\alpha^4}{8}
+\frac{8\,\alpha^5}{3\,\pi}+\alpha^6\ln\alpha\\[1ex]
&+&\left(\frac{7}{\pi^2}\,\zeta(3)-\frac{2}{\pi^2}-\frac{33}{16}\right)
\alpha^6+O(\alpha^7)\ ,\end{eqnarray*}where $\zeta(z)$ is the Riemann zeta
function $$\zeta(z)=\sum_{k=1}^\infty k^{-z}\ ,\quad\Re z>1\ ,$$ which
coincides up to order $O(\alpha^4)$ with the improved lower bound on the
ground-state energy;\item for $n_{\rm r}=1$, $\ell=0$:\begin{eqnarray*}
\frac{E(n_{\rm r}=1,\ell=0)}{m}&=&1-\frac{\alpha^2}{8}-\frac{13\,\alpha^4}{128}
+\frac{\alpha^5}{3\,\pi}+\frac{\alpha^6\ln\alpha}{8}\\[1ex]
&+&\left(\frac{7}{8\,\pi^2}\,\zeta(3)-\frac{1}{4\,\pi^2}-\frac{197}{1024}
-\frac{\ln 2}{8}\right)\alpha^6+O(\alpha^7)\ ;\end{eqnarray*}\item for
$n_{\rm r}=0$, $\ell=1$, where, remarkably, there are no contributions of the
order $O(\alpha^5)$ or $O(\alpha^6\ln\alpha)$: $$\frac{E(n_{\rm
r}=0,\ell=1)}{m}=1-\frac{\alpha^2}{8}-
\frac{7\,\alpha^4}{384}+\frac{727\,\alpha^6}{82944}+O(\alpha^7)\ .$$
\end{itemize} In any case, even for the Coulomb potential~(\ref{Eq:CP}) the
eigenvalues of the semirelativistic Hamiltonian~(\ref{Eq:SRH}) are not known
exactly!

\section{EXACT ANALYTIC UPPER BOUNDS ON ENERGY LEVELS}

In view of the (admittedly rather unsatisfactory) state-of-the-art mentioned
above, let us try to~find, at least, exact upper bounds on the eigenvalues of
the Hamiltonian (\ref{Eq:SRH}) by entirely analytic methods. Clearly, the
derivation of upper bounds on the eigenvalues of some operator only makes
sense if this operator is bounded from below. Accordingly, let us assume that
the (otherwise arbitrary) interaction potential $V$ is such that the
Hamiltonian $H$ is bounded from below. For the {\bf Coulomb
potential}~(\ref{Eq:CP}), the required semi-boundedness of the spectrum of
the Hamiltonian $H$ has been rigorously proven \cite{Herbst}.

\subsection{Minimum--Maximum Principle}Beyond doubt, the theoretical
foundation of (as well as the primary tool for) the derivation of rigorous
upper bounds on the eigenvalues of some self-adjoint operator is the
well-known ``min--max~principle.'' There are several equivalent formulations
of this theorem. For practical purposes,~the most convenient one is the
following:\begin{itemize}\item Let $H$ be a self-adjoint operator bounded
from below.\item Let $E_k$, $k=0,1,2,\dots$, denote the eigenvalues of $H$,
$$H|\chi_k\rangle=E_k|\chi_k\rangle\ ,\quad k=0,1,2,\dots\ ,$$ ordered
according to $$E_0\le E_1\le E_2\le\dots\ .$$\item Let $D_d$ be some
$d$-dimensional subspace of the domain of~$H$.\end{itemize} Then the $k$th
eigenvalue $E_k$ (counting multiplicity) of $H$ satisfies the inequality
$$\fbox{$E_k\le\displaystyle\sup_{|\psi\rangle\in D_{k+1}}
\frac{\langle\psi|H|\psi\rangle}{\langle\psi|\psi\rangle}$}\quad\mbox{for}\
k=0,1,2,\dots\ .$$

\subsection{Operator Inequalities}Needless to say, our goal must be to
replace the problematic kinetic-energy square-root operator~in~the
semirelativistic Hamiltonian $H$ by some more tractable operator. One
(obvious) way to achieve this~is to use the min--max principle in order to
compare eigenvalues of operators:\begin{itemize}\item Assume the validity of
a generic operator inequality of the form $$H\le{\cal O}\ .$$
Then\begin{eqnarray*}E_k&\equiv&
\frac{\langle\chi_k|H|\chi_k\rangle}{\langle\chi_k|\chi_k\rangle}\\[1ex]
&\le&\sup_{|\psi\rangle\in D_{k+1}}
\frac{\langle\psi|H|\psi\rangle}{\langle\psi|\psi\rangle}\qquad\mbox{(by the
min--max principle)}\\[1ex]&\le&\sup_{|\psi\rangle\in
D_{k+1}}\frac{\langle\psi|{\cal
O}|\psi\rangle}{\langle\psi|\psi\rangle}\qquad\mbox{(by the operator
inequality)}\ .\end{eqnarray*}\item Assume that the $(k+1)$-dimensional
subspace $D_{k+1}$ in these inequalities is spanned by the~first $k+1$
eigenvectors~of the operator ${\cal O}$, that is, by precisely those
eigenvectors of~${\cal O}$ that correspond to the first $k+1$ eigenvalues
$\widehat E_0,\widehat E_1,\dots,\widehat E_k$ of ${\cal O}$ if all
eigenvalues of ${\cal O}$ are ordered according~to $$\widehat E_0\le\widehat
E_1\le\widehat E_2\le\dots\ .$$ Then $$\sup_{|\psi\rangle\in
D_{k+1}}\frac{\langle\psi|{\cal
O}|\psi\rangle}{\langle\psi|\psi\rangle}=\widehat E_k\ .$$\end{itemize}
Consequently, every eigenvalue $E_k$ of $H$ is bounded from above by a
corresponding eigenvalue $\widehat E_k$~of~${\cal O}$:
$$\fbox{$E_k\le\widehat E_k$}\ .$$

\subsection[]{The ``Schr\"odinger'' Bound \cite{Lucha96a}}The simplest of all
upper bounds on the eigenvalues of the semirelativistic Hamiltonian $H$ of
Eq.~(\ref{Eq:SRH})~is the one involving the corresponding ``Schr\"odinger''
Hamiltonian. It may be found by taking advantage of the positivity of the
square of the (since $T$ is self-adjoint, obviously self-adjoint) operator
$T-m$:\begin{eqnarray*}
0&\le&(T-m)^2\\[1ex]&=&T^2+m^2-2\,m\,T\\[1ex]&\equiv&{\bf
p}^2+2\,m^2-2\,m\,T\ .\end{eqnarray*}Assuming $m$ to be positive, this may be
converted into an operator inequality for the kinetic energy~$T$, $$T\le
m+\frac{{\bf p}^2}{2\,m}\ ,$$ which, in turn, entails an operator inequality
for the (generic, since it involves an arbitrary interaction potential $V$)
Hamiltonian $H$ in the spinless Salpeter equation \cite{Lucha96a}: $$H\le
H_{\rm S}\ ,$$ with the ``Schr\"odinger'' Hamiltonian $$H_{\rm
S}=m+\frac{{\bf p}^2}{2\,m}+V\ .$$

For the {\bf Coulomb potential} (\ref{Eq:CP}), the energy eigenvalues
corresponding to the above Schr\"odinger Hamiltonian read $$E_{{\rm
S},n}=m\left(1-\frac{\alpha^2}{2\,n^2}\right),$$ where the total quantum
number $n$ is given in terms of both radial and orbital angular-momentum
quantum numbers $n_{\rm r}$ and $\ell$, respectively, by $$n=n_{\rm
r}+\ell+1\ ,\quad n_{\rm r}=0,1,2,\dots\ ,\quad\ell=0,1,2,\dots\ .$$

\subsection[]{A ``Squared'' Bound \cite{Lucha96a}}In order to improve the
above bound, one might be tempted to consider the square of the Hamiltonian
$H$: $$Q\equiv H^2=T^2+V^2+T\,V+V\,T\ .$$ The eigenvalue equation for this
squared Hamiltonian $Q$ will, of course, be solved by exactly the~same set of
eigenvectors $|\chi_k\rangle$ as the one for the original Hamiltonian $H$
with, however, the squares $E_k^2$~of~the corresponding energy eigenvalues
$E_k$ of $H$ as the eigenvalues of $Q$: $$Q|\chi_k\rangle=E_k^2|\chi_k\rangle\
,\quad k=0,1,2,\dots\ .$$ The positivity of the square of the self-adjoint
operator $T-m-V$,\begin{eqnarray*}
0&\le&(T-m-V)^2\\[1ex]&=&T^2+m^2+V^2-2\,m\,T+2\,m\,V-T\,V-V\,T\ ,
\end{eqnarray*}and the (because of the positivity of the operator ${\bf p}^2$
obviously valid) relation $$0\le m\le T$$ yield some operator inequality for
the anticommutator $T\,V+V\,T$ of kinetic energy $T$ and interaction
potential $V$ showing up in the squared Hamiltonian $Q$:
\begin{eqnarray*}T\,V+V\,T&\le&T^2+m^2+V^2-2\,m\,T+2\,m\,V\\[1ex]
&\equiv&{\bf p}^2+2\,m^2+V^2-2\,m\,T+2\,m\,V\\[1ex] &\le&{\bf
p}^2+V^2+2\,m\,V\ .\end{eqnarray*}Insertion of this intermediate result
implies an operator inequality for the squared Hamiltonian
$Q$~\cite{Lucha96a}: $$Q\le R\ ,$$ with the operator $$R\equiv 2\,{\bf
p}^2+m^2+2\,V^2+2\,m\,V\ .$$ According to the min--max principle, the squares
of the energy eigenvalues $E_k$ of the spinless Salpeter equation are
therefore bounded from above by the corresponding eigenvalues ${\cal
E}_{R,k}$ of the operator $R$: $$E_k^2\le{\cal E}_{R,k}$$ or
$$E_k\le\sqrt{{\cal E}_{R,k}}\ .$$

Only for the {\bf Coulomb potential} (\ref{Eq:CP}), the above operator $R$ is
of exactly the same structure~as~the ``Schr\"odinger'' Hamiltonian $H_{\rm
S}$, with, however, some ``effective'' orbital angular momentum quantum
number $L$ to be determined from the relation
$$L\,(L+1)=\ell\,(\ell+1)+\alpha^2\ ,$$ with the result
$$L=\frac{\sqrt{1+4\left[\ell\,(\ell+1)+\alpha^2\right]}-1}{2}\
,\quad\ell=0,1,2,\dots\ .$$ The eigenvalues ${\cal E}_R$ of the operator $R$
may thus be found by replacing in the Coulomb eigenvalues~$E_{{\rm S},n}$ the
angular momentum quantum number $\ell$ by the effective angular momentum
quantum number $L$: $${\cal
E}_{R,N}=m^2\left(1-\frac{\alpha^2}{2\,N^2}\right),$$ with the ``effective''
total quantum number $N$ defined by $$N=n_{\rm r}+L+1\ ,\quad n_{\rm
r}=0,1,2,\dots\ .$$ Unfortunately, these ``squared'' bounds lie above and are
thus worse than the ``Schr\"odinger'' bounds.

\subsection[]{A Straightforward Generalization \cite{Lucha96a}}The
``Schr\"odinger'' bounds may be improved by generalizing a little bit the
line of argument leading~to their derivation. The positivity of the square of
the (obviously self-adjoint) operator $T-\mu$, where~$\mu$~is an arbitrary
real parameter (with the dimension of mass),
\begin{eqnarray*}0&\le&(T-\mu)^2\\[1ex]&=&T^2+\mu^2-2\,\mu\,T\\[1ex]&\equiv&
{\bf p}^2+m^2+\mu^2-2\,\mu\,T\ ,\end{eqnarray*}implies a set of operator
inequalities for the kinetic energy $T$:\footnote{\normalsize\ See also
Ref.~\cite{Martin}.} $$T\le\frac{{\bf p}^2+m^2+\mu^2}{2\,\mu}\quad\mbox{for
all}\ \mu>0\ .$$ This immediately translates into a set of operator
inequalities for the semirelativistic Hamiltonian~$H$ \cite{Lucha96a}:
$$H\le\widehat H_{\rm S}(\mu)\quad\mbox{for all}\ \mu>0\ ,$$ with the
Schr\"odinger-like Hamiltonian $$\widehat H_{\rm S}(\mu)=\frac{{\bf
p}^2+m^2+\mu^2}{2\,\mu}+V\ .$$ Invoking again the min--max principle, any
energy eigenvalue $E_k$ of $H$ is thus bounded from above by the
corresponding eigenvalue $\widehat E_{{\rm S},k}(\mu)$ of the
Schr\"odinger-like Hamiltonian $\widehat H_{\rm S}(\mu)$ for given
positive~$\mu$, $$E_k\le\widehat E_{{\rm S},k}(\mu)\quad\mbox{for all}\
\mu>0\ ,$$ and---as a (trivial) consequence---also by the minimum of all
these Schr\"odinger-like upper bounds~\cite{Lucha96a}:
$$\fbox{$E_k\le\displaystyle\min_{\mu>0}\widehat E_{{\rm S},k}(\mu)$}\ .$$

\newpage

For the {\bf Coulomb potential} (\ref{Eq:CP}), it is trivial to write down
the corresponding energy eigenvalues: $$\widehat E_{{\rm S},n}(\mu)=
\frac{1}{2\,\mu}\left[m^2+\mu^2\left(1-\frac{\alpha^2}{n^2}\right)\right],$$
with the total quantum number $$n=n_{\rm r}+\ell+1\ .$$ Minimizing $\widehat
E_{{\rm S},n}(\mu)$ with respect to the parameter $\mu$ yields \cite{Lucha96a}
\begin{equation}\fbox{$\displaystyle\min_{\mu>0}\widehat E_{{\rm
S},n}(\mu)=m\,\sqrt{1-\frac{\alpha^2}{n^2}}$}\quad\mbox{for all}\
\alpha\le\alpha_{\rm c}\ .\label{Eq:GUB}\end{equation} These bounds
\begin{itemize}\item hold for all values $\alpha\le\alpha_{\rm c}$ of the
Coulomb coupling constant $\alpha$ and arbitrary levels of excitation,
and\item improve definitely the Schr\"odinger bounds, for any value of the
total (or ``principal'') quantum number $n$: $$\min_{\mu>0}\widehat E_{{\rm
S},n}(\mu)<E_{{\rm S},n}\quad\mbox{for}\ \alpha\neq 0\ .$$\end{itemize} (For
$\mu=m$ one necessarily recovers the Schr\"odinger case.) The {\bf
comparison} of the upper bounds~of Eq.~(\ref{Eq:GUB}) with their numerically
\cite{Raynal} or by a perturbative expansion \cite{Brambilla} obtained
counterparts shows~that\begin{enumerate}\item for the ground state, i.e.,
$n_{\rm r}=\ell=0$, hence $n=1$ (Table~\ref{Tab:Naub}), the relative error of
our analytical (operator) bound (in the considered range of the coupling
constant $\alpha$) is less than 4.5~$\%$, and

\vspace*{-2.5ex}

\begin{table}[h]\caption[]{Numerical \cite{Raynal} vs. analytical upper
bounds on the ground-state energy $E_0$}\label{Tab:Naub}
\begin{center}\begin{tabular}{lll}\hline\hline&&\\[-1.5ex]
\multicolumn{1}{c}{$\alpha$}&\multicolumn{2}{c}{Upper Bound on
$\displaystyle\frac{E_0}{m}$}\\[2ex]
\multicolumn{1}{c}{}&\multicolumn{1}{c}{$\qquad$Numerical$\qquad$}&
\multicolumn{1}{c}{$\qquad$Analytical$\qquad$}\\[1ex]\hline&&\\[-1.5ex]
$\quad$0.0155522$\quad$&$\qquad$0.9998791&$\qquad$0.9998791\\
$\quad$0.1425460$\quad$&$\qquad$0.989613&$\qquad$0.989788\\
$\quad$0.2599358$\quad$&$\qquad$0.96364&$\qquad$0.96563\\
$\quad$0.3566678$\quad$&$\qquad$0.92673&$\qquad$0.93423\\
$\quad$0.4359255$\quad$&$\qquad$0.88139&$\qquad$0.8998\\
$\quad$0.5$\quad$&$\qquad$0.82910&$\qquad$0.86603\\[1ex]
\hline\hline\end{tabular}\end{center}\end{table}

\vspace*{-2.5ex}

\item for the level $n_{\rm r}=0$, $\ell=1$ and thus $n=2$
(Table~\ref{Tab:Pceeaub}) the relative error of our analytical~(operator)
bound (in the considered range of the coupling constant $\alpha$) is less
than 0.1~$\%$.

\vspace*{-2.5ex}

\begin{table}[h]\caption[]{Perturbatively computed \cite{Brambilla} energy
eigenvalues vs. analytical upper bounds for $n_{\rm
r}=0$,~$\ell=1$}\label{Tab:Pceeaub}
\begin{center}\begin{tabular}{lll}\hline\hline&&\\[-1.5ex]
\multicolumn{1}{c}{$\alpha$}&
\multicolumn{2}{c}{$\displaystyle\frac{E(n_{\rm r}=0,\ell=1)}{m}$}\\[2ex]
\multicolumn{1}{c}{}&\multicolumn{1}{c}{\hspace{1ex}Perturbation
Theory\hspace{1ex}}&\multicolumn{1}{c}{\hspace{1ex}Analytical
Bound\hspace{1ex}}\\[1ex]\hline&&\\[-1.5ex]
$\quad$0.0155522$\quad$&$\qquad$0.999969765&$\qquad$0.999969766\\
$\quad$0.1425460$\quad$&$\qquad$0.997452&$\qquad$0.997457\\
$\quad$0.2599358$\quad$&$\qquad$0.99147&$\qquad$0.99152\\
$\quad$0.3566678$\quad$&$\qquad$0.9838&$\qquad$0.9840\\
$\quad$0.4359255$\quad$&$\qquad$0.975&$\qquad$0.976\\
$\quad$0.5$\quad$&$\qquad$0.967&$\qquad$0.9682\\[1ex]
\hline\hline\end{tabular}\end{center}\end{table}
\end{enumerate}

\vspace*{-2.5ex}

\subsection{Rayleigh--Ritz Variational Technique}An {\bf immediate
consequence of the min--max principle} is the famous Rayleigh--Ritz
technique:\begin{itemize}\item Restrict the operator $H$ to the subspace
$D_d$ by orthogonal projection $P$ onto $D_d$: $$\left.\widehat H\equiv
H\right|_{D_d}:=P\,H\,P\ .$$\item Let $\widehat E_k$, $k=0,1,\dots,d-1$,
denote all $d$ eigenvalues of this restricted operator $\widehat H$,
$$\widehat H|\widehat\chi_k\rangle=\widehat E_k|\widehat\chi_k\rangle\ ,\quad
k=0,1,\dots,d-1\ ,$$ ordered according to $$\widehat E_0\le\widehat
E_1\le\dots\le\widehat E_{d-1}\ .$$\end{itemize} Then the $k$th eigenvalue
$E_k$ (counting multiplicity) of $H$ satisfies the inequality
$$\fbox{$E_k\le\widehat E_k$}\ ,\quad k=0,1,\dots,d-1\ .$$

Assume now that this $d$-dimensional subspace $D_d$ is spanned by some set of
$d$ linearly independent basis vectors $|\psi_k\rangle$, $k = 0,1,\dots,d-1$:
Then the set of eigenvalues $\widehat E$ may immediately be determined~by
diagonalizing the $d\times d$ matrix $$\left(\langle\psi_i|\widehat
H|\psi_j\rangle\right),\quad i,j=0,1,\dots,d-1\ ,$$ i.e., as the $d$ roots of
the characteristic equation $$\det\left(\langle\psi_i|\widehat
H|\psi_j\rangle-\widehat E\,\langle\psi_i|\psi_j\rangle\right)=0\ ,\quad
i,j=0,1,\dots,d-1\ .$$ (This becomes clear by expanding any eigenvector of
the restricted operator $\widehat H$ in terms of the basis vectors
$|\psi_k\rangle$, $k=0,1,\dots,d-1$, of the subspace $D_d$.)

\subsection[]{Variational Bound for the Ground State
\cite{Lucha94,Lucha96a}}In order to derive, as a first application of the
Rayleigh--Ritz variational technique, an upper bound~on the ground-state
energy eigenvalue $E_0$, it is sufficient to focus one's interest to the case
$k=0$ (i.e.,~to consider an only one-dimensional subspace). In this
particular case, the min--max principle reduces~to
$$E_0\le\frac{\langle\psi|H|\psi\rangle}{\langle\psi|\psi\rangle}\ .$$ (This
inequality simply states that the ground-state energy $E_0$ is less than or
equal to any expectation value of the Hamiltonian $H$.) However, one is
certainly entitled to consider simultaneously even~sets~of one-dimensional
trial spaces and to compute optimized upper bounds by the following {\bf
prescription} (or ``recipe''):\begin{enumerate}\item Choose a suitable set of
trial states $\{|\psi_\lambda\rangle\}$ (with elements distinguished from
each other by~some variational parameter $\lambda$). Each of these trial
states serves to span a one-dimensional trial~space.\item Calculate the
expectation values of the Hamiltonian $H$ with respect to these trial states
$|\psi_\lambda\rangle$:
$$E(\lambda)\equiv\frac{\langle\psi_\lambda|H|\psi_\lambda\rangle}
{\langle\psi_\lambda|\psi_\lambda\rangle}\ .$$\item Determine (from the first
derivative of $E(\lambda)$ with respect to $\lambda$) that value
$\lambda_{\rm min}$ of the variational parameter $\lambda$ which minimizes
$E(\lambda)$.\item Compute $E(\lambda_{\rm min})$ (that is, the minimal
expectation value of $H$ in the Hilbert-space subsector~of the chosen trial
states $\{|\psi_\lambda\rangle\}$).\end{enumerate} When going through these
steps, your reward will be an optimized upper bound for the ground~state:
$$E_0\le E(\lambda_{\rm min})\ .$$

\newpage

In order to get rid of the troublesome square-root operator in the
Hamiltonian $H$, one may adopt a trivial (nevertheless fundamental)
inequality for the expectation values of a self-adjoint but otherwise
arbitrary operator ${\cal O}={\cal O}^\dagger$ and its square, taken with
respect to any arbitrary Hilbert-space state~$|\psi\rangle$ in the domain of
this operator: $$\langle\psi|{\cal O}|\psi\rangle^2\le\langle\psi|{\cal
O}^2|\psi\rangle\,\langle\psi|\psi\rangle$$ and thus
$$\fbox{$\displaystyle\frac{|\langle\psi|{\cal
O}|\psi\rangle|}{\langle\psi|\psi\rangle}\le\sqrt{\frac{\langle\psi|{\cal
O}^2|\psi\rangle}{\langle\psi|\psi\rangle}}$}\ .$$ It should prove to be
advantageous to consider the above inequality for the kinetic-energy
operator~$T$:
$$\frac{|\langle\psi|T|\psi\rangle|}{\langle\psi|\psi\rangle}\le
\sqrt{\frac{\langle\psi|T^2|\psi\rangle}{\langle\psi|\psi\rangle}}\ .$$ Apply
this to the semirelativistic Hamiltonian $H$: \begin{eqnarray*}
E_0&\le&\frac{\langle\psi|H|\psi\rangle}{\langle\psi|\psi\rangle}
\qquad\mbox{(by the min--max principle)}\\[1ex]
&=&\frac{\langle\psi|T+V|\psi\rangle}{\langle\psi|\psi\rangle}\\[1ex]
&\le&\sqrt{\frac{\langle\psi|T^2|\psi\rangle}{\langle\psi|\psi\rangle}}
+\frac{\langle\psi|V|\psi\rangle}{\langle\psi|\psi\rangle}\qquad\mbox{(by the
above inequality)}\\[1ex]&\equiv&\sqrt{\frac{\langle\psi|{\bf
p}^2|\psi\rangle}{\langle\psi|\psi\rangle}+m^2}
+\frac{\langle\psi|V|\psi\rangle}{\langle\psi|\psi\rangle}\ .\end{eqnarray*}
Note, with due satisfaction, that now only the expectation values of ${\bf
p}^2$ and of $V$ have to be evaluated.

For the {\bf Coulomb potential} (\ref{Eq:CP}), maybe the first choice for the
coordinate-space representation~of trial vectors which comes to one's mind
are the well-known (normalized) hydrogen-like trial functions
$$\psi_\lambda({\bf
x})=\sqrt{\displaystyle\frac{\lambda^3}{\pi}}\,\exp(-\lambda\,r)\
,\quad\lambda>0\ .$$ This choice of trial states yields for the expectation
values\begin{itemize}\item of the square of the momentum $\bf p$:
$$\frac{\left\langle\psi_\lambda\left|{\bf
p}^2\right|\psi_\lambda\right\rangle}{\langle\psi_\lambda|\psi_\lambda\rangle}
=\lambda^2\ ,$$\item of the inverse of the radial coordinate $r$:
$$\frac{\left\langle\psi_\lambda\left|r^{-1}\right|\psi_\lambda\right\rangle}
{\langle\psi_\lambda|\psi_\lambda\rangle}=\lambda\ .$$\end{itemize} Insertion
of these expectation values into the above boundary expression results in the
one-parameter set of upper bounds
$$E_0\le\sqrt{\lambda^2+m^2}-\alpha\,\lambda\quad\mbox{for all}\ \lambda>0\
,$$ with the absolute minimum $$\fbox{$E_0\le m\,\sqrt{1-\alpha^2}$}\ ;$$
this is identical to the previously found generalized operator bound for
$n=1$. For $\alpha\neq 0$ this bound~is lower and thus better than the
Schr\"odinger bound on the ground-state energy (which is characterized by the
quantum numbers $n_{\rm r}=\ell=0$ and thus $n=1$): $$E_{{\rm
S},0}=m\left(1-\frac{\alpha^2}{2}\right).$$ One may conclude that the
variational technique yields indeed improved upper bounds on the energy
levels.

\subsection[]{Energy Levels at the Critical Coupling Constant
\cite{Lucha96b}}For the {\bf Coulomb potential} (\ref{Eq:CP}), it might be of
interest to derive upper bounds on the energy~levels~at the critical coupling
constant $\alpha_{\rm c}$, particularly, to improve the upper bound on the
ground-state~energy. A still rather simple-minded but very useful choice for
the basis vectors $|\psi_k\rangle$ (labelled by some positive integer $k$,
i.e., $k=0,1,2,\dots$) of the $d$-dimensional trial space $D_d$ is given
\begin{itemize}\item in configuration-space representation by $$\psi_k(r)=
\sqrt{\frac{(2\,m)^{2\,k+2\,\beta+1}}{4\pi\,\Gamma(2\,k+2\,\beta+1)}}\,
r^{k+\beta-1}\exp(-m\,r)\ ,\quad r\equiv|{\bf x}|\ ,\quad\beta\ge 0\ ,\quad
m>0\ ,$$\item in momentum-space representation by $$\widetilde\psi_k(p)=
\sqrt{\frac{(2\,m)^{2\,k+2\,\beta+1}}{2\pi^2\,\Gamma(2\,k+2\,\beta+1)}}\,
\Gamma(k+\beta+1)\,\frac{\sin\left[(k+\beta+1)\arctan\displaystyle\frac{p}{m}
\right]}{p\,(p^2+m^2)^{(k+\beta+1)/2}}\ ,\quad p\equiv|{\bf p}|\
,\quad\beta\ge 0\ .$$\end{itemize}

Here, the parameter $\beta$ allows, for a given value of the coupling
constant $\alpha$, of the total cancellation of the divergent contributions
to the expectation values of,\begin{itemize}\item on the one hand, the
kinetic-energy operator $T$ for large momenta $p$ and,\item on the other
hand, the Coulomb interaction-potential operator $V_{\rm C}(r)$ at small
distances $r$:\end{itemize} $$\beta=\beta(\alpha)\ .$$ The parameter $\beta$
is implicitly determined as a function of the coupling constant~$\alpha$, for
instance,~by~the relation\footnote{\normalsize\ Choosing equidistant values
for the parameter $\beta$, this relation then yields the curious
values~of~the Coulomb coupling constant $\alpha$ adopted in
Tables~\ref{Tab:Nolaub}, \ref{Tab:Naub}, and
\ref{Tab:Pceeaub}.}~\cite{Durand,Nickisch}
$$\alpha=\beta\cot\left(\frac{\pi}{2}\,\beta\right)\quad\mbox{for the ground
state}\ .$$ Hence, the critical coupling constant, $\alpha_{\rm c}$, is
approached in the limit of vanishing $\beta$, that is, for $\beta\to 0$. For
the present choice of trial states, the above-mentioned singularities arise
only in matrix elements taken with respect to the ground state
$|\psi_0\rangle$, i.e., in the matrix elements
$\langle\psi_0|T|\psi_0\rangle$ and $\langle\psi_0|V_{\rm
C}(r)|\psi_0\rangle$. For arbitrary parameter value $\beta$, the above choice
of basis vectors $|\psi_k\rangle$ implies for the matrix elements
\begin{itemize}\item of the kinetic-energy operator $T$
\begin{eqnarray*}\langle\psi_i|T|\psi_j\rangle
&=&\displaystyle\frac{2^{i+j+2\,\beta+1}\,m}{\pi}\,
\displaystyle\frac{\Gamma(i+\beta+1)\,\Gamma(j+\beta+1)}
{\sqrt{\Gamma(2\,i+2\,\beta+1)\,\Gamma(2\,j+2\,\beta+1)}}\\[3ex]
&\times&\displaystyle\int\limits_0^\infty{\rm d}y\,
\displaystyle\frac{\cos[(i-j)\arctan y]-\cos[(i+j+2\,\beta+2)\arctan y]}
{(1+y^2)^{(i+j+2\,\beta+1)/2}}\ ,\end{eqnarray*}which has to be evaluated
with the help of the expansion $$\cos(N\arctan
y)=\displaystyle\frac{1}{(1+y^2)^{N/2}}\,\displaystyle\sum_{n=0}^N
\left(\begin{array}{c}N\\n\end{array}\right)\cos\left(\frac{n\,\pi}{2}\right)
y^n\quad\mbox{for}\ N=0,1,2,\dots\ ;$$\item of the Coulomb
interaction-potential operator $V=V_{\rm C}(r)$
$$\langle\psi_i|V|\psi_j\rangle=-\frac{2\,m\,\alpha\,\Gamma(i+j+2\,\beta)}
{\sqrt{\Gamma(2\,i+2\,\beta+1)\,\Gamma(2\,j+2\,\beta+1)}}\ ;$$\item and for
the projections of the basis states $|\psi_k\rangle$ onto each other
$$\langle\psi_i|\psi_j\rangle=\frac{\Gamma(i+j+2\,\beta+1)}
{\sqrt{\Gamma(2\,i+2\,\beta+1)\,\Gamma(2\,j+2\,\beta+1)}}\ .$$\end{itemize}

For notational simplicity, define a dimensionless energy eigenvalue
$\varepsilon$ by extracting overall~factors: $$\widehat
E=:\frac{2}{\pi}\,m\,\varepsilon\ .$$ The resulting characteristic equation
for the case of relevance here, viz., $\beta=0$, is typically of the~form
$$\det\left(\begin{array}{ccc}4\ln 2-2-\varepsilon&
\displaystyle\frac{\sqrt{2}}{3}-\displaystyle\frac{\varepsilon}{\sqrt{2}}&
\cdots\\[2ex]
\displaystyle\frac{\sqrt{2}}{3}-\displaystyle\frac{\varepsilon}{\sqrt{2}}&
\displaystyle\frac{17}{15}-\varepsilon&\cdots\\[2ex]
\vdots&\vdots&\ddots\end{array}\right)=0\ .$$ The roots of the characteristic
equation (which may be calculated algebraically up to $d=4$)
then~read,\begin{itemize}\item for $d=1$, $$\varepsilon=2\,(2\ln 2-1)\ ,$$
which entails, as upper bound for the ground state, $$\frac{\widehat
E_0}{m}=0.4918\dots\ ;$$\item for $d=2$,
$$\varepsilon=\frac{1}{15}\left(60\ln 2-23\pm\sqrt{(60\ln 2)^2-4800\ln
2+1649}\right),$$ which entails, as upper bound for the ground state,
$$\frac{\widehat E_0}{m}=0.484288\dots\ ;$$\item while, for $d=4$, a lengthy
expression for the roots implies, as upper bound for the ground~state,
$$\fbox{$\displaystyle\frac{\widehat E_0}{m}=0.4842564\dots$}\
.$$\end{itemize} Interestingly, already for $d=2$ our (analytical) bound lies
well within the numerically obtained range $$0.4825\le\frac{E_0}{m}\le
0.4842910\quad\mbox{for}\ \alpha=\alpha_{\rm c}\
.$$

\subsection{Generalized Laguerre Basis}

Quite generally, upper bounds on eigenvalues may be improved by (suitably)
modifying the trial~space $D_d$, by enlarging it to higher dimensions $d$ or
by spanning it by a more sophisticated set of basis~states. A---rather
popular---choice of trial functions is based on the generalized Laguerre
polynomials $L_k^{(\gamma)}(x)$ (for the parameter $\gamma$): these are
orthogonal polynomials\begin{itemize}\item defined by the power series
$$L_k^{(\gamma)}(x)=\sum_{t=0}^k\,(-1)^t\left(\begin{array}{c}k+\gamma\\k-t
\end{array}\right)\frac{x^t}{t!}$$\item and normalized, with the weight
function $x^\gamma\exp(-x)$, according to $$\int\limits_0^\infty{\rm
d}x\,x^\gamma\exp(-x)\,L_k^{(\gamma)}(x)\,
L_{k'}^{(\gamma)}(x)=\frac{\Gamma(\gamma+k+1)}{k!}\,\delta_{kk'}\ .$$
\end{itemize}In order to construct trial vectors $|\psi\rangle$ of the
subspace $D_d$, corresponding to states with orbital angular momentum $\ell$
and its projection $m$, introduce two variational parameters, namely:
\begin{itemize}\item $\mu$ (with dimension of mass), and\item $\beta$
(dimensionless).\end{itemize}

\newpage

\no The chosen set of ``Laguerre'' trial states $|\psi_k\rangle$ is then
defined by the configuration-space representation $$\psi_{k,\ell m}({\bf
x})=\sqrt{\frac{(2\,\mu)^{2\,\ell+2\,\beta+1}\,k!}
{\Gamma(2\,\ell+2\,\beta+k+1)}}\,r^{\ell+\beta-1}\exp(-\mu\,r)\,
L_k^{(2\,\ell+2\,\beta)}(2\,\mu\,r)\,{\cal Y}_{\ell m}(\Omega_{\bf x})\ ,$$
with the spherical harmonics ${\cal Y}_{\ell m}(\Omega)$ for angular momentum
$\ell$ and its projection $m$, depending on~the solid angle $\Omega$ and
orthonormalized according to $$\int{\rm d}\Omega\,{\cal Y}^\ast_{\ell
m}(\Omega)\,{\cal Y}_{\ell'm'}(\Omega)=\delta_{\ell\ell'}\,\delta_{mm'}\ .$$
The requirement of normalizability of the (Hilbert-space) trial states
$|\psi_k\rangle$ imposes on the parameters $\mu$ and $\beta$ the constraints
$$\mu>0\ ,\quad\beta>-\frac{1}{2}\ .$$ In this case, the configuration-space
trial function $\psi_{k,\ell m}({\bf x})$ satisfies the orthonormalization
condition $$\int{\rm d}^3x\,\psi_{k,\ell m}^\ast({\bf
x)}\,\psi_{k',\ell'm'}({\bf
x)}=\delta_{kk'}\,\delta_{\ell\ell'}\,\delta_{mm'}\ .$$ The momentum-space
representation of these trial states $|\psi_k\rangle$ (obtained by Fourier
transformation)~is\begin{eqnarray*}\widetilde\psi_{k,\ell m}({\bf p})
&=&\sqrt{\frac{(2\,\mu)^{2\,\ell+2\,\beta+1}\,k!}
{\Gamma(2\,\ell+2\,\beta+k+1)}}\,\frac{(-{\rm i})^\ell\,|{\bf
p}|^\ell}{2^{\ell+1/2}\,\Gamma\left(\ell+\frac{3}{2}\right)}\,\sum_{t=0}^k\,
\frac{(-1)^t}{t!}\left(\begin{array}{c}k+2\,\ell+2\,\beta\\
k-t\end{array}\right)\\[1ex]
&\times&\frac{\Gamma(2\,\ell+\beta+t+2)\,(2\,\mu)^t}{({\bf
p}^2+\mu^2)^{(2\,\ell+\beta+t+2)/2}}\,
F\left(\frac{2\,\ell+\beta+t+2}{2},-\frac{\beta+t}{2};\ell+\frac{3}{2};
\frac{{\bf p}^2}{{\bf p}^2+\mu^2}\right){\cal Y}_{\ell m}(\Omega_{\bf p})\
,\end{eqnarray*}with the hypergeometric series $F$, defined by
$$F(u,v;w;z)=\frac{\Gamma(w)}{\Gamma(u)\,\Gamma(v)}\,\sum_{n=0}^\infty\,
\frac{\Gamma(u+n)\,\Gamma(v+n)}{\Gamma(w+n)}\,\frac{z^n}{n!}\ .$$ Then
trivially, the momentum-space trial function $\widetilde\psi_{k,\ell m}({\bf
p})$ satisfies the orthonormalization condition $$\int{\rm
d}^3p\,\widetilde\psi_{k,\ell m}^\ast({\bf
p)}\,\widetilde\psi_{k',\ell'm'}({\bf
p)}=\delta_{kk'}\,\delta_{\ell\ell'}\,\delta_{mm'}\ .$$

\subsection[]{Power-Law Potentials \cite{Lucha97}}In order to be fairly
general, consider an interaction potential of power-law form:
$$V(r)=\sum_na_n\,r^{b_n}\ ,\quad r\equiv|{\bf x}|\ ,$$ where the sets of
(otherwise arbitrary) real constants $a_n$ and $b_n$ are only constrained by
the necessary boundedness from below of the Hamiltonian:
$$b_n\ge-1\quad\mbox{if}\quad a_n<0\ .$$ The matrix elements of this
power-law potential $V(|{\bf x}|)$ are easily worked
out:\begin{eqnarray*}\langle\psi_i|V|\psi_j\rangle
&=&\sqrt{\frac{i!\,j!}{\Gamma(2\,\ell+2\,\beta+i+1)\,
\Gamma(2\,\ell+2\,\beta+j+1)}}\,\sum_n\,\frac{a_n}{(2\,\mu)^{b_n}}\,
\sum_{t=0}^i\,\sum_{s=0}^j\,\frac{(-1)^{t+s}}{t!\,s!}\\[1ex]
&\times&\left(\begin{array}{c}i+2\,\ell+2\,\beta\\i-t\end{array}\right)
\left(\begin{array}{c}j+2\,\ell+2\,\beta\\j-s\end{array}\right)
\Gamma(2\,\ell+2\,\beta+b_n+t+s+1)\ .\end{eqnarray*}From this expression, the
potential matrix $V\equiv(\langle\psi_i|V|\psi_j\rangle)$ may easily be
written down explicitly: if\begin{itemize}\item considering, for instance,
only radial excitations ($\ell=0$) and\item choosing for the variational
parameter $\beta$ the value $\beta=1$,\end{itemize} one finds
$$V=\frac{1}{6}\,\sum_n\,\frac{a_n}{(2\,\mu)^{b_n}}\,\Gamma(3+b_n)
\left(\begin{array}{ccc}3&-\sqrt{3}\,b_n&\cdots\\[1ex]
-\sqrt{3}\,b_n&3+b_n+b_n^2&\cdots\\[1ex]
\vdots&\vdots&\ddots\end{array}\right).$$

\subsection[]{Analytically Evaluable Special Cases \cite{Lucha97}}Compared
with the above calculation, an evaluation of the kinetic-energy matrix
elements $\langle\psi_i|T|\psi_j\rangle$ is certainly more delicate. However,
there are several situations which allow for an analytical treatment.

\subsubsection{\bf Orbital Excitations} Restricting to the case $i=j=0$ but
allowing for arbitrary values of the orbital angular momentum~$\ell$ means to
consider pure orbital excitations. Merely for definiteness, fix, for the
moment, the variational parameter $\beta$ to the value $\beta=1$. In this
case, the matrix elements of the power-law potential operator simplify to
$$\langle\psi_0|V|\psi_0\rangle=\frac{1}{\Gamma(2\,\ell+3)}\,\sum_n\,
\frac{a_n}{(2\,\mu)^{b_n}}\,\Gamma(2\,\ell+b_n+3)\ ,$$ while a rather
straightforward calculation yields for the matrix elements of the
kinetic-energy operator
$$\langle\psi_0|T|\psi_0\rangle=\frac{4^{\ell+2}\,[\Gamma(\ell+2)]^2}
{\sqrt{\pi}\,\Gamma\left(2\,\ell+\frac{7}{2}\right)}\,\mu\,
F\left(-\frac{1}{2},\ell+2;2\,\ell+\frac{7}{2};1-\frac{m^2}{\mu^2}\right).$$
Aiming at analytical results, one should clearly try to get rid of the
hypergeometric series $F$. There~are several possibilities to do so:
\begin{itemize}\item In the ultrarelativistic limit (realized for vanishing
particle mass, $m=0$), by use of the relation
$$F(u,v;w;1)=\frac{\Gamma(w)\,\Gamma(w-u-v)}{\Gamma(w-u)\,\Gamma(w-v)}\quad
\mbox{for}\ w\ne 0,-1,-2,\dots\ ,\ \Re(w-u-v)>0\ ,$$ the above kinetic-energy
matrix element simplifies to
$$\langle\psi_0|T|\psi_0\rangle=\frac{2\,[\Gamma(\ell+2)]^2}
{\Gamma\left(\ell+\frac{3}{2}\right)\Gamma\left(\ell+\frac{5}{2}\right)}\,\mu\
.$$ For instance, for the {\bf linear potential} $$\quad V(r)=a\,r\ ,\quad
a>0\ ,$$ minimizing the expectation value $\langle\psi_0|H|\psi_0\rangle$
with respect to the variational parameter $\mu$ leads~to the minimal upper
bound $$\fbox{$\displaystyle\min_{\mu>0}\langle\psi_0|H|\psi_0\rangle=2\,
\Gamma(\ell+2)\,\sqrt{\frac{(2\,\ell+3)\,a}{\Gamma\left(\ell+\frac{3}{2}\right)
\Gamma\left(\ell+\frac{5}{2}\right)}}=2\,\sqrt{2\,a}\,\frac{\Gamma(\ell+2)}
{\Gamma\left(\ell+\frac{3}{2}\right)}$}\ .$$ In the limit of large orbital
angular momenta ($\ell\to\infty$), this optimized upper bound reduces to
$$\lim_{\ell\to\infty}\left(\min_{\mu>0}\langle\psi_0|H|\psi_0\rangle\right)^2
=8\,a\left(\ell+\frac{5}{4}\right).$$ Consequently, it describes linear Regge
trajectories:\footnote{\normalsize\ The square of the exact optimized bound
is already almost perfectly linear. Its deviation from the asymptotic
linearity is monotonously decreasing for increasing angular momentum $\ell$.
The maximum relative deviation (occuring clearly at the point $\ell=0$) is
$1-5\,\pi/16=1.8\ \%$.} $$[E(\ell)]^2\propto\ell\ ,$$ which is in striking
accordance with all other findings based on different considerations
\cite{Kang,Lucha91}.\item Another possibility---which will not be followed
any longer---is to fix the variational parameter~$\mu$ to the value $\mu=m$
(which reduces the last argument of the hypergeometric series $F$ to 0)
and~to take advantage of the relation $$F(u,v;w;0)=1\ ,$$ in order to obtain
the kinetic-energy matrix element
$$\langle\psi_0|T|\psi_0\rangle=\frac{4^{\ell+2}\,[\Gamma(\ell+2)]^2}
{\sqrt{\pi}\,\Gamma\left(2\,\ell+\frac{7}{2}\right)}\,m\ .$$\end{itemize}

\subsubsection{Radial Excitations} Focusing one's interest exclusively to
states with vanishing orbital angular momentum $\ell$ ($\ell=0$) means to
consider only radial excitations. The matrix elements of the kinetic-energy
operator then~typically take the form\begin{eqnarray*}
\langle\psi_i|T|\psi_j\rangle&=&\sqrt{\frac{i!\,j!}{\Gamma(2\,\beta+i+1)\,
\Gamma(2\,\beta+j+1)}}\,\frac{4^{\beta+1}}{2\pi}\,\mu\,
\sum_{t=0}^i\,\sum_{s=0}^j\,\frac{(-2)^{t+s}}{t!\,s!}\\[1ex]
&\times&\left(\begin{array}{c}i+2\,\beta\\i-t\end{array}\right)
\left(\begin{array}{c}j+2\,\beta\\j-s\end{array}\right)
\Gamma(\beta+t+1)\,\Gamma(\beta+s+1)\,I_{ts}\ ,\end{eqnarray*}with some
integral $I_{ts}$ which under certain circumstances may be calculated
analytically. For instance, for $\mu=m$, and for $2\,\beta$ integer and thus
(remembering the above normalizability constraint $2\,\beta>-1$)
non-negative, i.e., $2\,\beta=0,1,2,\dots$, one easily
finds\begin{eqnarray*}I_{ts}&=&\frac{1}{2}
\left[\Gamma\left(\frac{2\,\beta+t+s+|t-s|+1}{2}\right)\right]^{-1}
\sum_{n=0}^{|t-s|}\left(\begin{array}{c}|t-s|\\n\end{array}\right)\\[1ex]
&\times& \Gamma\left(\frac{n+1}{2}\right)
\Gamma\left(\frac{2\,\beta+t+s+|t-s|-n}{2}\right)
\cos\left(\frac{n\,\pi}{2}\right)\\[1ex]
&-&\frac{1}{2}\left[\Gamma\left(2\,\beta+t+s+\frac{3}{2}\right)\right]^{-1}
\sum_{n=0}^{2\,\beta+t+s+2}
\left(\begin{array}{c}2\,\beta+t+s+2\\n\end{array}\right)\\[1ex]
&\times&\Gamma\left(\frac{n+1}{2}\right)
\Gamma\left(2\,\beta+t+s+1-\frac{n}{2}\right)
\cos\left(\frac{n\,\pi}{2}\right).\end{eqnarray*}With this result, the
kinetic-energy matrix $T\equiv(\langle\psi_i|T|\psi_j\rangle)$ reads, for
variational parameter $\beta$ fixed~to $\beta=1$, explicitly:
$$T=\frac{64}{15\,\pi}\,m\left(\begin{array}{ccc}1&
\displaystyle\frac{\sqrt{3}}{7}&\cdots\\[2ex]
\displaystyle\frac{\sqrt{3}}{7}&\displaystyle\frac{11}{9}&\cdots\\[2ex]
\vdots&\vdots&\ddots\end{array}\right).$$In this way, analytic expressions
for the matrix elements $\langle\psi_i|H|\psi_j\rangle$ of~the
semirelativistic Hamiltonian $H$ with power-law potential may be found.
\begin{itemize}\item Up to a trial-space dimension $d=4$ these matrix
elements are, at least in principle, algebraically accessible.\item For
larger trial-space dimension, i.e., $d>4$, the energy matrix
$(\langle\psi_i|H|\psi_j\rangle)$ must be diagonalized numerically, without,
however, the necessity to invoke (time-consuming) integration
procedures.\end{itemize}

In order to estimate the quality of all the upper bounds derived, consider,
for the following reason, the {\bf harmonic-oscillator potential} $$\quad
V(r)=\omega\,r^2\ ,\quad\omega>0\ .$$ In momentum space the operator $r^2$ is
represented by the Laplacian with respect to the momentum~${\bf p}$,
$$r^2\longrightarrow\Delta_{\bf p}\ ,$$ while the kinetic energy, nonlocal in
configuration space, is represented by a multiplication operator.
Consequently, the semirelativistic Hamiltonian $H$ in momentum-space
representation is equivalent~to a nonrelativistic Hamiltonian with some
effective (square-root) interaction potential. The equation~of motion
resulting in this way is then solvable with one of the numerous numerical
procedures designed for the treatment of the nonrelativistic Schr\"odinger
equation. For the {\bf comparison} (Table~\ref{Tab:Uboel}) of these Laguerre
bounds with results of a numerical solution of the (``momentum-space,'' in
the above sense) Schr\"odinger equation, we employ some standard set of
parameter values \cite{Lucha92}: $m=2\;\mbox{GeV}$, $\omega=2\;\mbox{GeV}^3$
(as well as the choice $\mu=m$, $\beta=1$); we observe a rapid convergence of
our upper bounds towards~the exact energy eigenvalues.

\vspace*{-2.5ex}

\begin{table}[ht]\caption{Upper bounds (in units of GeV) on energy levels
(labelled in usual spectroscopic notation) for a $d\times d$ energy matrix.
(Analytically obtained results in italics.)}\label{Tab:Uboel}
\begin{center}\begin{tabular}{ccccc}\hline\hline\\[-1.5ex]
\ \ State\ \ &\ \ $1\times 1$\ \ &\ \ $2\times 2$\ \ &\ \ $25\times 25$\ \ &\
\ Schr\"odinger\ \ \\[1ex]
\hline\\[-1.5ex]
1S&{\it 4.2162}&{\it 3.9276}&3.8249&3.8249\\
2S&---&{\it 8.1085}&5.7911&5.7911\\
3S&---&---&7.4829&7.4823\\
4S&---&---&9.0215&9.0075\\[1ex]
\hline\hline\end{tabular}\end{center}\end{table}

\vspace*{-2.5ex}

\section{SUMMARY}The min--max principle and related simple theorems allow to
derive rigorous statements on the energy levels predicted by a bound-state
wave equation like the spinless Salpeter equation. In principle, these
methods should work for all physical situations which may be formulated as an
eigenvalue problem. If one is not satisfied with the outcome of this
procedure, or if one is interested in the corresponding~wave functions, one
has to actually solve the equation of motion by some numerical approximation
method. The most efficient among these is perhaps the ``semianalytical matrix
method'' proposed in Ref.~\cite{Lucha92nam}. Nevertheless, there still remain
questions of the significance \cite{Lucha92} of too naive relativistic
improvements of some nonrelativistic potential models for the description of
hadrons as bound states of quarks \cite{Lucha91prep}.

\end{document}